\documentclass[%
 reprint,
 amsmath,amssymb,
 aps,
]{revtex4-2}

\usepackage{graphicx}
\usepackage{dcolumn}
\usepackage{bm}
\usepackage{color}
\usepackage{amsfonts}
\usepackage{dsfont}
\usepackage{amsmath}
\usepackage{amssymb}
\usepackage{changes}
\usepackage{comment}
\usepackage{wrapfig}
\usepackage{hyperref}
\usepackage{tcolorbox}

\newtheorem{result}{Result}

 \newcommand{\ket}[1]{|#1\rangle}
 \newcommand{\bra}[1]{\langle #1|}
 \newcommand{\ketbra}[1]{|#1\rangle\!\langle #1|}

\newcommand{\M}{\mathcal{M}}

\newcommand{\eye}{\mathds{1}}

\newcommand{\Tr}{\text{Tr}}
\newcommand{\D}{\text{d}}

\begin{document}


\title{Verification of Continuous-Variable Quantum Memories}

\author{Paolo Abiuso}\email{paolo.abiuso@oeaw.ac.at}
\affiliation{Institute for Quantum Optics and Quantum Information - IQOQI Vienna,
Austrian Academy of Sciences, Boltzmanngasse 3, A-1090 Vienna, Austria}
\affiliation{ICFO – Institut de Ciencies Fotoniques, The Barcelona Institute of Science and Technology, 08860 Castelldefels (Barcelona), Spain}


\begin{abstract}
    A proper quantum memory is argued to consist in a quantum channel which cannot be simulated with a measurement followed by classical information storage and a final state preparation, i.e. with an entanglement breaking (EB) channel.
    The verification of quantum memories (non-EB channels) is a task in which an honest user wants to test the quantum memory of an untrusted, remote provider. This task is inherently suited for the class of protocols with trusted quantum inputs, sometimes called measurement-device-independent (MDI) protocols.  
    
    Here, we study the MDI certification of non-EB channels in continuous variable (CV) systems. We provide a simple witness based on adversarial metrology, and describe an experimentally friendly protocol that can be used to verify all non Gaussian incompatibility breaking quantum memories. Our results can be tested with current technology and can be applied to test other devices resulting in non-EB channels, such as CV quantum transducers and transmission lines.
\end{abstract}

\maketitle

\section{Introduction}
Quantum memories are a fundamental hardware component in present and future quantum communication technologies.
For example, a quantum memory is necessary to perform quantum teleportation~\cite{bennett1993teleporting,vaidman1994teleportation,braunstein1998teleportation}, entanglement swapping, and to build quantum repeaters~\cite{briegel1998quantum}. 
Schemes of measurement-based quantum computing also rely on hardware that can can preserve entangled quantum states~\cite{nielsen2003quantum}.
At the same time, in recent years highly efficient quantum memories have been built, both for finite-dimensional systems~\cite{zhao2009millisecond,specht2011single-atom}, as well as continuous-variable (CV) bosonic systems~\cite{julsgaard2004experimental,hedges2010efficient}, reaching more than 50\% efficiency on the storage timescale of $\mu s$ to $ms$.
It is therefore crucial to develop tests that certify the proper functioning of a quantum memory with minimal assumptions. This is relevant both for calibration and benchmarking purposes, as well as for adversarial settings, in which part of the hardware is owned by an untrusted party.
For example, near-future quantum computation and communication devices are most likely to be held by big tech ``quantum providers", while users with limited technological capacity can remotely use their services.

What is a good quantum memory? Arguably, a memory should correspond to a channel $\M$ that is as close as possible to the identity $\mathcal{M}\sim\eye$. However it is clear that, for instance, a unitary channel $U$ corresponds to a perfect memory, as any state can then be recovered by applying the inverse $U^\dagger$. Rather, from an operational point of view, the main property a quantum memory should satisfy, is that of not being replaceable by the simple storage of classical data, collected from the input state, followed by a state preparation at the desired time $\tau$. This class of \emph{measure and prepare} channels coincides with entanglement breaking (EB) channels~\cite{horodecki2003entanglement}, which can be always decomposed as
\begin{align}
\label{eq:EBchannel}
    \mathcal{E}[\rho]=\int \D a\; \rho^{(a)}\Tr[N^{a}\rho]\;,
\end{align}
$N^{(a)}$ being a positive operator-valued measure (POVM), to each outcome of which is associated the preparation of the state $\rho^{(a)}$.
The verification of quantum memories thus formally translates to the \emph{certification of non entanglement breaking channels} (non-EB). 
In particular, in~\cite{rosset2018resource} the authors introduced a formal framework for quantum memory certification with minimal assumptions and constructively provided a protocol that can witness \emph{any finite dimensional non-EB memory}, that only requires an honest user to produce fiduciary input states and send them to the untrusted memory provider, which outputs classical data. 
Such protocol was verified in recent experiments~\cite{mao2020experimentally,graffitti2020measurement,yu2021measurement}.

However, an important slice of quantum technologies relies on the standard bosonic, continuous variable (CV) quantum information framework~\cite{braunstein2005quantum,ferraro2005gaussian,adesso2014continuous}, for which previous results in the literature cannot be applied.
In this work, we consider the certification of quantum memories in such regime. 
We introduce an experimentally-friendly certification protocol for CV quantum memories that is based on an adversarial displacement estimation, and can be used to certify a vast class of non-EB CV quantum memories (namely, all non Gaussian incompatibility breaking channels~\cite{heinosaari2015breaking}) using Gaussian operations and measurements only.

\section{Quantum channel verification in finite dimensions}
The idea of quantum channel verification has been explored in various contexts:
in~\cite{pusey15quantum}, Pusey analyzed the semi-device-independent certification of non-EB channels, in a 1-mode setup, showing that the only nontrivial scenario is the one in which the preparation device is trusted, but not the measurement device, the so called measurement-device-independent (MDI) scenario~\cite{buscemi2012all,branciard2013measurement,abiuso2021measurement}, and finding that in such case EB channels can be witnesses when inducing incompatible measurements~\cite{pusey15quantum,heinosaari2015incompatibility};
an extensive analysis of the correlations generated by EB and other classes of channels 
is given in Ref.~\cite{ku2022quantifying}; 
a fully device-independent (DI) approach was proposed by Sekatski \emph{et al.}~\cite{sekatski2018certifying} (see also the recent~\cite{neves2023experimental,sekatski2023towards}) to certify  quantum channels (up to isometries) based on self-testing; 
Dall'Arno \emph{et al.} also worked on DI tests of quantum channels~\cite{dallarno2017DI_channels} and measurements~\cite{dallarno2017DI_measurements}, obtaining characterizations in specific small-dimensional cases.

All the protocols proposed in the mentioned works however suffer from the same crucial assumption: the possibility of deciding whether the channel $\mathcal{M}$ has been implemented or not. In memory verification scenarios where the honest user Alice is technology-limited and can only process classical data received from the untrusted provider Eve, it is clear that this assumption cannot be safely taken, as Eve could perform all their operations and measurements at the very beginning of the protocol, without using the memory at all.
Rosset \emph{et al.}~\cite{rosset2018resource} solved this issue by considering two input modes with a delay between them, without the necessity of using entangled sources.
In their protocol, Alice can send quantum inputs to Eve, who claims to have a good quantum memory, i.e. a non-EB channel  $\mathcal{M}:\mathcal{S}(\mathbb{C}^d)\rightarrow \mathcal{S}(\mathbb{C}^d)$ -- $\mathcal{S}(\mathbb{C}^d)$ being the set of trace-one positive operators on $\mathbb{C}^d$ --
that can store quantum states $\rho$ of dimension $d$ for some time $\tau$, before returning $\mathcal{M}[\rho]$. Alice sends to Eve a state $\rho$ at time $t=0$, and another state $\phi$ at time $t=\tau$. Eve can perform any measurement on $ \rho$ and $\phi$, but given the time delay, she is forced to use $\mathcal{M}[\rho]$ in order to perform a joint measurement. 
If Eve performs projections on the maximally entangled state, then Alice can certify if $\mathcal{M}$ is non-EB, merely from the statistics $P(b|\rho,\phi)$. Let's say, in fact, that Eve's POVM is 
\begin{align}
    M^{b=0} =\ket{\psi^+}\bra{\psi^+}\;,\quad
    M^{b=1} &=\eye - \ket{\psi^+}\bra{\psi^+}\;,
\end{align}
where $\ket{\psi^+}:=\sum_{i=1}^d\frac{1}{\sqrt{d}}\ket{ii}$ is the maximally entangled state in dimension $d$.
Then, one has
\begin{align}
\label{eq:Pb_ross}
    P(b=0|\rho,\phi)=\frac{1}{d}\Tr[\phi^\intercal\mathcal{M}[\rho]]\;.
\end{align}
That is, this choice of measurement formally connects the output statistics to the tomography of $\mathcal{M}$ itself. To certify the non-EB property, consider an entanglement witness $W$ such that
\begin{align}
    \Tr[W (\eye\otimes\mathcal{M})[\psi^+]] > 0\;,\;\;
    \Tr[W (\eye\otimes\mathcal{E})[\psi^+]] \leq 0\;,
\end{align}
for any EB channel $\mathcal{E}$. That is, $W$ is an entanglement witness for the Choi-Jamiolkowski state of $\mathcal{M}$.
Given a decomposition 
\begin{align}
\label{eq:W_deco}
    W=\sum_{ij} c_{ij} \rho_i^{\intercal}\otimes\phi^\intercal_j\;,
\end{align}
one has, combining~\eqref{eq:Pb_ross} and~\eqref{eq:W_deco},
\begin{align}
    \sum_{ij} c_{ij} P(b=0|\rho_i,\phi_j)
    & =\Tr[W (\eye\otimes\mathcal{M})[\psi^+]]> 0\;.
\end{align}
Instead, if the memory is EB, it can be simulated by a measure-and-prepare channel~\eqref{eq:EBchannel}, and regardless of the chosen measurement by Eve, the resulting statistics is of the form
$
    P(b=0|\rho_i,\phi_j)=\int \D a \Tr[(N^{a}\otimes M^{b=0|a})\rho_i\otimes\phi_j]
$,
where $N^{a}$ and $M^{b=0|a}$ are POVM elements. In such case simple algebra shows that the same combination $\sum c_{ij}P(b=0|\rho_i,\phi_j)$ is proportional to the average value of $W$ applied on a separable state, and therefore it is negative semidefinite~\cite{rosset2018resource}.

It is clear however that this construction has few drawbacks: it cannot be practically scaled to large dimensions, as it involves the projection on maximally entangled states, and the associated probabilities scale as $1/d$. This makes it particularly unsuitable for the case of continuous variables, both at the theoretical and practical level~\footnote{In the context of continuous variables, naively one could substitute the maximally entangled state $\ket{\psi^+}$ with two-mode squeezed states 
\begin{align*}
    \ket{\Psi^{(r)}}:=\sqrt{1-\tanh{r}^2}\sum_{i=0}^{\infty} \tanh{r}^i\ket{ii}\;,
\end{align*}
and perform the same protocol proposed by the authors of~\cite{rosset2018resource}. However, the measurement projecting on such states is not of easy implementation~\cite{braunstein2005quantum}(cf. also Sup.Mat. of~\cite{abiuso2021measurement}), and for any amount of squeezing $r$ there exist non-EB channels $\M_r$ such that $\M_r[\Psi^{(r)}]$ is separable.}. Moreover, one needs to know in advance the specific witness $W$ associated to $\M$. 
We overcome these limitations, and inspired by the case of MDI CV entanglement detection~\cite{abiuso2021measurement}, we provide in the following an experimentally-friendly protocol that can detect a large class of CV quantum memories, relying on a single witness, based on phase-space displacement metrology~\cite{genoni2013optimal}.

\section{Experimentally-friendly MDI witness for CV quantum memories}
Here we outline our protocol for the (MDI) verification of continuous variables quantum memories.

\begin{tcolorbox}[title=A note on the choice of units for CV states.]
In this paper we use standard units for bosonic operators and quadratures, in which
\begin{align*}
    \hat{x}=\frac{\hat{a}+\hat{a}^\dagger}{\sqrt{2}}\;,\quad 
    \hat{p}= \frac{\hat{a}-\hat{a}^\dagger}{\sqrt{2}i}\;,
\end{align*}
where $a$, $a^\dagger$ are the canonical annihilation and creation operators.
It follows that $[\hat{x},\hat{p}]=i$, and the uncertainty principle reads ${\rm Var}[\hat{p}]{\rm Var}[\hat{x}]\geq \frac{1}{4}$. Coherent states are defined as
\begin{align*}
    \ket{\alpha}=e^{i \sqrt{2}\alpha_p \hat{x}-i\sqrt{2}\alpha_x\hat{p}}\ket{0}=e^{\alpha \hat{a}^{\dagger}-\alpha^*\hat{a}}\ket{0}\;,
\end{align*}
where $\alpha=\alpha_x+ i\alpha_p$.
In these units, the quadratures' variance on coherent states, including the vacuum, is equal to $\frac{1}{2}$.
\end{tcolorbox}

\begin{figure*}
    \centering
    \includegraphics[width=0.8\textwidth]{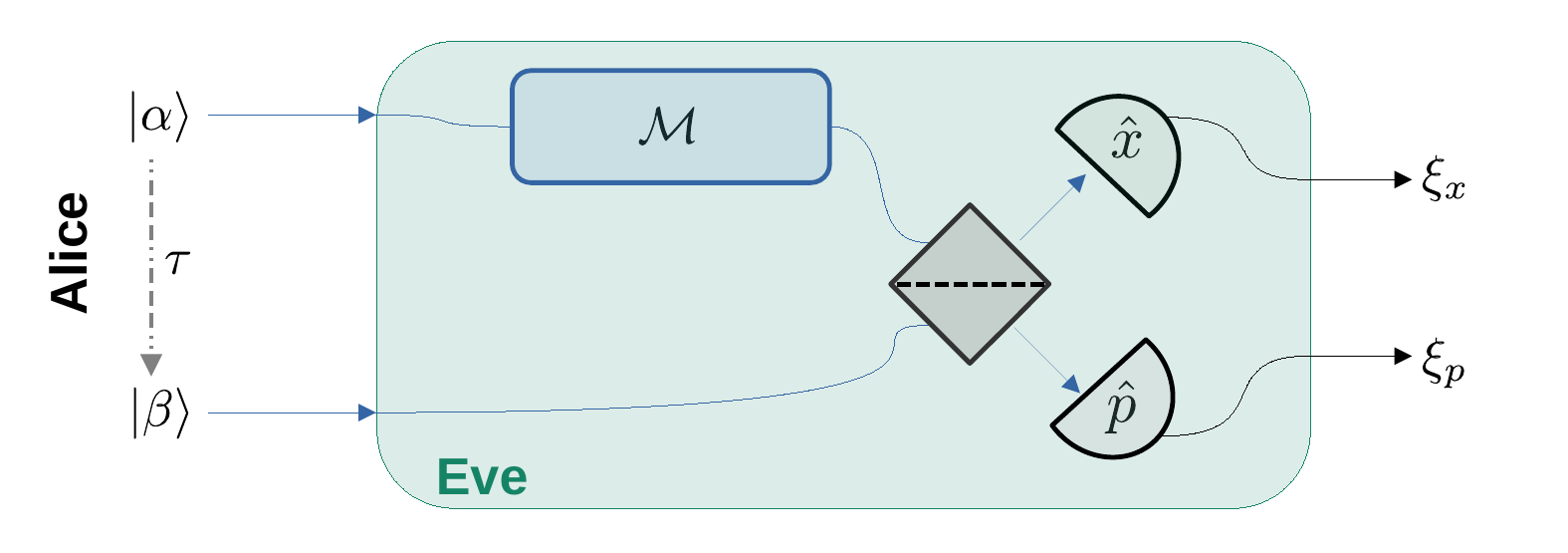}
    \caption{Alice sends random coherent states $\ket{\alpha}$, and, with a delay $\tau$, $\ket{\beta}$. Untrusted Eve, after storing $\ket{\alpha}$ in the memory $\mathcal{M}$, measures jointly the two states and outputs $\{\xi_x$,$\xi_p$\}, trying to estimate the values of $\alpha_x+\beta_x$  and $\alpha_p-\beta_p$ with small error, in order to minimize the average value of the witness~\eqref{eq:score}. If the memory is perfect ($\mathcal{M}=\eye$), the optimal strategy consists in mixing the two signals in a balanced beamsplitter, followed by homodyne measurements, resulting in
    $\hat{\xi}_x\equiv\hat{x}_E=\frac{\hat{x}_\alpha+\hat{x}_\beta}{\sqrt{2}}$ and $\hat{\xi}_p\equiv\hat{p}_E=\frac{\hat{p}_\alpha-\hat{p}_\beta}{\sqrt{2}}$.
    If the memory is entanglement breaking, the minimum error ``doubles", assuming the sampling distributions of $\alpha$ and $\beta$ to be wide enough (cf. Result~\ref{res1}). We only assume the faithful characterization of the input coherent states, while the inequality~\eqref{eq:EB_bound_main} is valid regardless of the (EB) memory and measurement apparatuses operated by the untrusted Eve. 
    When Eve has a non-gIB memory, the inequality~\eqref{eq:EB_bound_main} can be violated by performing the same measurements, possibly appending Gaussian channels before ($\mathcal{G}_1$) and after ($\mathcal{G}_2$) the quantum storage $\M$ (cf. Result~\ref{res2}).}
    \label{fig:setup}
\end{figure*}

\paragraph*{Informal explanation of the protocol.}
\label{sec:witness_friendly}
The setup is simple, as illustrated in  Figure~\ref{fig:setup}. Alice sends a random coherent state $\ket{\alpha}$, and after some time $\tau$ (the nominal time of the memory), $\ket{\beta}$. If the memory is perfect, $\ket{\alpha}$ is left unperturbed, and Eve can measure the quadratures
\begin{align}
\hat{x}_E:=\frac{\hat{x}_\alpha+\hat{x}_\beta}{\sqrt{2}}\;,\quad
\hat{p}_E:=\frac{\hat{p}_\alpha-\hat{p}_\beta}{\sqrt{2}}\;,
\end{align}
via a beam-splitter and homodyne detection. This corresponds to
getting an estimate of $\alpha_x+\beta_x$ and $\alpha_p-\beta_p$ with one vacuum noise each, that is
\begin{align}
    \langle\hat{x}_E\rangle &=\alpha_x+\beta_x \;, &   \langle\hat{p}_E\rangle &=\alpha_p-\beta_p\;,\\
    {\rm Var}[\hat{x}_E] &=\frac{1}{2}\;, & {\rm Var}[\hat{p}_E] &=\frac{1}{2}\;.
    \label{eq:XE_PE_Var}
\end{align}
We turn this metrological task into a witness: Alice inputs random coherent states labelled by $\alpha$ and $\beta$, and we ask Eve to return two outputs $\xi_x$ and $\xi_p$, computing the average value of the resulting score
\begin{align}
\label{eq:score}
 \mathcal{W}:= \left(\xi_x - (\alpha_x+\beta_x)\right)^2 +
  \left(\xi_p - (\alpha_p-\beta_p)\right)^2\;.
\end{align}
Eve is therefore presented with a metrology problem, i.e. estimating $\alpha_x+\beta_x$ and $\alpha_p-\beta_p$ at the same time, with the smallest possible error.
In case of perfect memory, $\mathcal{M}\equiv\eye$, she can perform the above protocol and the average value will be equal to $1$, due to~\eqref{eq:XE_PE_Var},
\begin{align}
\label{eq:id_score}
    \langle\mathcal{W}\rangle|_{\mathcal{M}=\eye}=1 \;.
\end{align}
Now, if instead the channel $\mathcal{M}$ is entanglement breaking, this means that it can be simulated with a measure-and-prepare channel~\eqref{eq:EBchannel}. Therefore, any estimation strategy by Eve can be simulated by one in which $\ket{\alpha}$ and $\ket{\beta}$ are measured separately. In such case, the estimation of $\alpha_x+\beta_x$ and $\alpha_p-\beta_p$ the above quantity will need to go through all the four values of $\alpha_x,\beta_x,\alpha_p,\beta_p$. Due to the incompatibility of measuring $\hat{x}$ and $\hat{p}$ simultaneously, the error gets doubled in this case.
That is, the optimal strategy, if $\mathcal{M}$ is entanglement breaking, becomes measuring $\frac{\hat{x}_\alpha}{\sqrt{2}}$ and $\frac{\hat{p}_\alpha}{\sqrt{2}}$ with a double homodyne measurement~\cite{genoni2013optimal}, and later the same for $\frac{\hat{x}_\beta}{\sqrt{2}}$ and $\frac{\hat{p}_\beta}{\sqrt{2}}$. By doing so, the average error increases, i.e. $\langle\mathcal{W}\rangle\sim\frac{4}{2}=2$. Following this reasoning, we thus expect that for an EB memory $\mathcal{M}$, 
$\langle\mathcal{W}\rangle|_{\mathcal{M}\in\text{EB}}\gtrsim 2 $.
This is the fundamental idea behind our memory witness. In Appendix~\ref{sec:EB_bound_proof} we formalize this idea by proving the following

\begin{result}
\label{res1}
Consider the above protocol in which uncorrelated random coherent states $\ket{\alpha}$,$\ket{\beta}$ are sent (with a delay between them) to Eve, which stores $\ket{\alpha}$ in their memory and is then able to perform any joint measurement on $\mathcal{M}[\ketbra{\alpha}]\otimes\ketbra{\beta}$. If $\mathcal{M}$ is entanglement breaking, the minimum value of $\langle\mathcal{W}\rangle$~\eqref{eq:score} is bounded by
\begin{align}
\label{eq:EB_bound_main}
    \langle\mathcal{W}\rangle \geq \frac{\sigma_\alpha^2}{1+\sigma_\alpha^2} +\frac{\sigma_\beta^2}{1+\sigma_\beta^2}\;.
\end{align}
Here $\sigma_{\alpha,\beta}$ correspond to the width of the distribution with which $\{\alpha,\beta\}$ are sampled, which we assume to be Gaussian for simplicity, i.e. $P(\alpha)=(\pi\sigma^2_\alpha)^{-1}{\rm Exp}[-|\alpha|^2/\sigma^2_\alpha]$ (this requirement can be relaxed, see App.~\ref{app:non-gaussian_priors}). 
\end{result}
The formalization of the above inequality relies on a Bayesian Cramér-Rao bound, which takes into account the input sampling distributions $P(\alpha)$ and $P(\beta)$~\cite{yuen1973multiple,genoni2013optimal}.
The meaning of the lower bound in Eq.~\eqref{eq:EB_bound_main} is the following: if the input distributions of $\alpha$ and $\beta$ have small variance, after a finite amount of rounds Eve can get an estimate of the input distribution, and start guessing using this additional information. In the limit of $\sigma_\alpha\rightarrow 0$ and $\sigma_\beta\rightarrow 0$ it is clear that Eve can simply output the combinations of average values $\xi_x=\langle{\alpha}_x\rangle+\langle{\beta}_x\rangle$, $\xi_p=\langle{\alpha}_p\rangle-\langle{\beta}_p\rangle$  and obtain a perfect score $\mathcal{W}\rightarrow 0$, at each round of the experiment. Viceversa, in the opposite limit of large $\sigma_{\alpha,\beta}$, the inputs become completely random, the optimal estimation strategy by Eve cannot be boosted by any prior knowledge of $P(\alpha,\beta)$, and the lower bound tends to $2$, which is obtained by measuring $\ket{\alpha}$ and $\ket{\beta}$ separately with double homodyne measurements.
The technical details are given in App.~\ref{sec:EB_bound_proof}.

\section{Violating the witness with good quantum memories}
Our main inequality~\eqref{eq:EB_bound_main} provides a bound on the metrological capabilities induced by CV EB  memories. Together with~\eqref{eq:id_score}, it is clear that a memory consisting in a channel sufficiently close to the identity $\eye$, can violate such bound and be witnessed as non-EB. It is therefore natural to ask how noise resistant is such witnessing, and more in general what is the largest class of (non-EB) channels that can be witnessed by the violation of~\eqref{eq:EB_bound_main}. We provide an answer for the case of Gaussian channels, via the following result, which is proven in Appendix~\ref{sec:gauss_app}.

\begin{result}
\label{res2}
Any 
memory $\mathcal{M}$ consisting in a Gaussian channel that is not Gaussian incompatibility breaking (gIB)~\cite{heinosaari2015breaking}, can be used to obtain a score~\eqref{eq:score} $\langle\mathcal{W}\rangle<2$, by appending Gaussian channels $\mathcal{G}_{1,2}$ to it and performing the above described protocol (cf. Fig.~\ref{fig:setup}) with $\mathcal{M'}\equiv\mathcal{G}_2\circ\M\circ\mathcal{G}_1$. By choosing $\sigma_{\alpha,\beta}$ large enough, this implies the violation of the bound~\eqref{eq:EB_bound_main} and certifies the memory $\M$ to be non-EB.
\end{result}
The channels $\mathcal{G}_{1,2}$ can be thought as pre- and post-processing that can only increase the EB-ness of the resulting memory~\footnote{Appending quantum channels is, in fact, a free transformation of channels in the resource theory of quantum memories~\cite{rosset2018resource}.}, but might be necessary in order to correct displacements/attenuations before performing the measurements (it is clear, for example, that if $\M=\mathcal{U}$ is a unitary channel, one can use $\mathcal{G}_1=\eye$ and $\mathcal{G}_2=\mathcal{U}^\dagger$ in order to have a perfect memory $\mathcal{M}'=\eye$).

In order to fully understand Result~\ref{res2}, one needs to clarify the involved set of quantum channels.
First, we limit ourselves to Gaussian channels: as in the standard literature, these are defined as those that preserve Gaussianity of the characteristic function of the state~\cite{adesso2014continuous,eisert2005gaussian,braunstein2005quantum}. Among these, Gaussian incompatibility breaking (gIB) channels are those that make all Gaussian observables jointly measurable, or \emph{compatible}, in the Heisenberg picture~\cite{heinosaari2015breaking}.
EB channels are known to break the incompatibility of any set of measurements~\cite{pusey15quantum,heinosaari2015incompatibility} (which is a direct consequence of Eq.~\eqref{eq:EBchannel}) and are therefore a subset of gIB channels, see Fig.~\ref{fig:eta_photonloss}. 
For a detailed discussion, see Appendix~\ref{app:IB_EB_etc}.

Finally, to decline our Results~\ref{res1} and~\ref{res2} to a an explicit example, we consider the paradigmatic case of a noisy memory affected by photon-loss noise.

\begin{figure}
    \centering
    \includegraphics[width=0.49\textwidth]{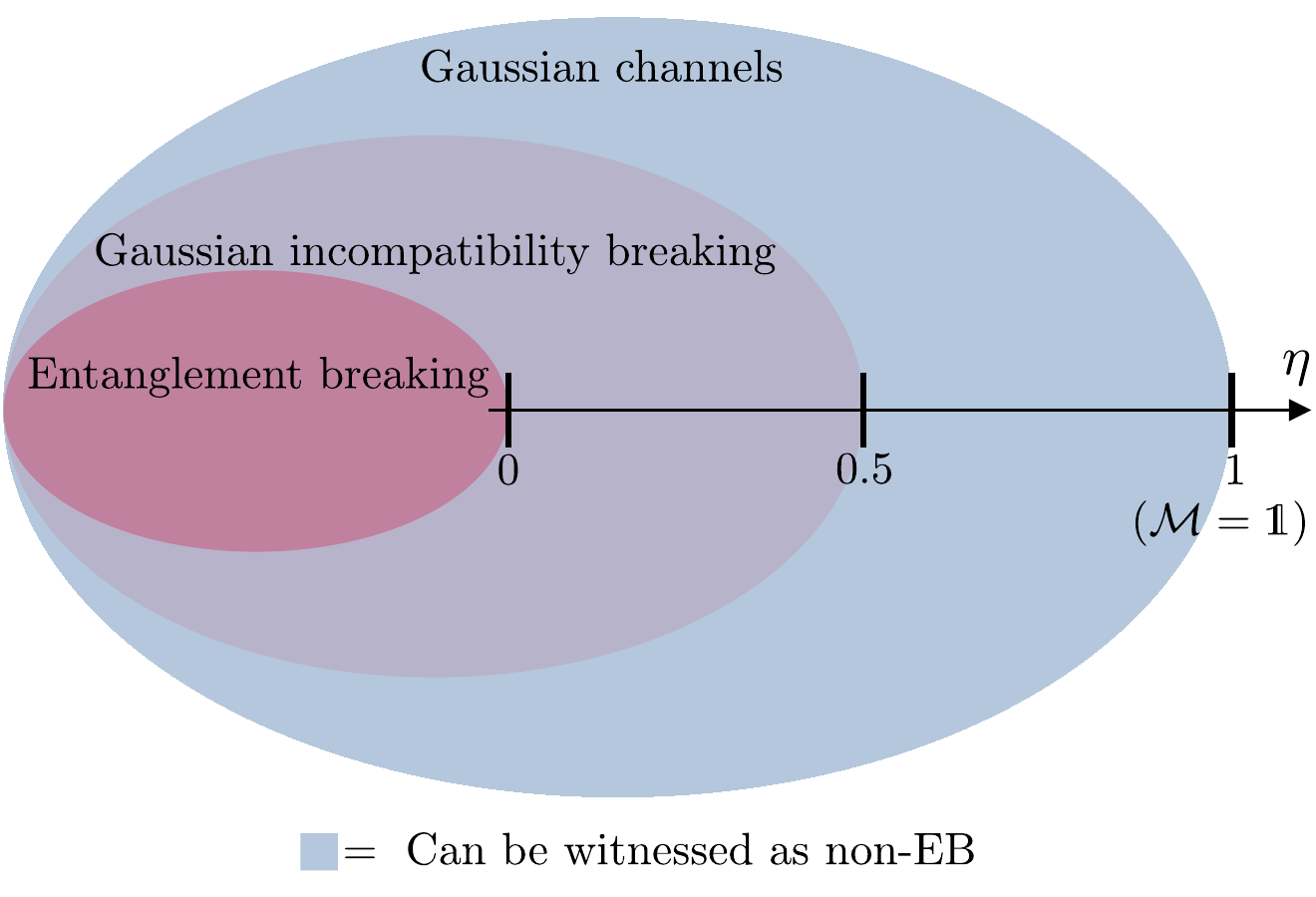}
    \caption{EB channels break the incompatibility of all measurements, and are therefore a subset of gIB channels~\cite{heinosaari2015breaking,heinosaari2015incompatibility}. All non-gIB channels can be witnessed with our protocol (Result~\ref{res2}).
    In case of a memory affected by photon losses, characterized by efficiency $\eta$: $\eta=0$ corresponds to a (EB) complete erasure of information; for $\eta\in (0,0.5]$ such channel is non-EB but is still gIB; for $\eta\in(0.5,1]$ the channel is non-gIB and can be witnessed by our protocol.}
    \label{fig:eta_photonloss}
\end{figure}

\subsection{Example: photon-loss}
Suppose the memory $\mathcal{M}$ is subject to simple photon-loss noise in mode $A$, which can be modelled as a beam splitter with transmissivity $0\leq\eta\leq 1$ (i.e. the memory efficiency). In the Heisenberg picture, this is described by
$
\hat{a}'_A=\sqrt{\eta}\hat{a}_A+\sqrt{1-\eta}\hat{a}^{(0)}_{A}\ , 
$
where $\hat{a}^{(0)}$ is a vacuum mode, and $\hat{a}_A$ is the original noiseless mode.
Then, a similar expression holds for the evolved quadratures $\hat{r}'$,
\begin{align}
\label{eq:PL_quadrature_heis}
 \hat{r}'_A=\sqrt{\eta}\hat{r}_A+\sqrt{1-\eta}\hat{r}^{(0)}_A\;, \quad r=x,p\;.
\end{align}
Then, to have an unbiased estimation of $\alpha_x+\beta_x$ and $\alpha_p-\beta_p$, one needs to apply an amplifier channel~\cite{caves1982quantum,weedbrook2012gaussian}, 
resulting in $\hat{r}''$ ($\nu\geq1$)
\begin{align}
\hat{r}''_A=\sqrt{\nu}\hat{r}'_A+\sqrt{\nu-1}\hat{r}'^{(0)}_A .\;
\end{align}
Before mixing the two modes $A$ and $B$, one needs to attenuate $\ket{\beta}$ in order to match the total attenuation of $\alpha$, i.e.
\begin{align}
\hat{r}'_B &=\sqrt{\eta\nu}\hat{r}_B+\sqrt{1-\eta\nu}\hat{r}^{(0)}_B\;, \quad r=x,p\;.
\end{align}
The final detection consists in mixing the $\hat{a}''_A$ and $\hat{a}'_B$ modes in the balanced beamsplitter, measure the two quadratures at the output ports, and finally apply a correction $(\eta\nu)^{-\frac{1}{2}}$
\begin{align}
\hat{\xi}_x:=\frac{\hat{x}''_A+\hat{x}'_B}{\sqrt{2\eta\nu}}\;,\quad
\hat{\xi}_p:=\frac{\hat{p}''_A-\hat{p}'_B}{\sqrt{2\eta\nu}}\;.
\end{align}
In this way, the estimation is unbiased, i.e. $\langle\hat{\xi}_x\rangle=\alpha_x+\beta_x$ and $\langle\hat{\xi}_p\rangle = \alpha_p-\beta_p$.
It follows that the resulting average value of the witness~\eqref{eq:score} is given by the variance of the estimators, which can be easily computed substituting the above equations, yielding
\begin{align}
\label{eq:PL_res}
    \langle\mathcal{W}\rangle(\eta,\nu)={\rm Var}[\hat{\xi}_x] + {\rm Var}[\hat{\xi}_p] = \frac{1}{\eta}\;.
\end{align}
Interestingly, in this case the value of $\langle\mathcal{W}\rangle$ does not depend on the amplification $\nu$, which can therefore be chosen at will. 
Comparing~\eqref{eq:PL_res} with the witnessing threshold $\langle\mathcal{W}\rangle<2$, we notice that for $\eta\in (0.5,1]$, the memory can be witnessed by our inequality, performing the protocol above. This is in accordance with Result~\ref{res2}, as $\eta=\frac{1}{2}$ is the Gaussian incompatibility breaking threshold of the photon-loss channel~\eqref{eq:PL_quadrature_heis} (see App.~\ref{app:gauss_channels}). as represented in Fig.~\ref{fig:eta_photonloss}.
Interestingly, the same threshold value for $\eta$ was found in~\cite{braunstein2000criteria} for the purpose of a inherently quantum teleportation, in a device-dependent scenario.

\section{Conclusion and comments}

In this work we studied the certification of CV quantum memories, and showed that it is possible to verify the quantumness of an untrusted remote CV memory by only assuming faithful description of input coherent states, i.e. in the MDI framework~\cite{rosset2018resource}. In particular, we presented a single inequality (Result~\ref{res1}) that cannot be violated by any EB channel, and at the same time can be used to detect a large class of non-EB Gaussian channels (Result~\ref{res2}), in a simple protocol that only requires homodyne measurements.
We notice that, as for the case of quantum key distribution~\cite{pirandola2015high-rate} and entanglement detection~\cite{abiuso2021measurement}, moving from the DI framework to MDI, allows practical protocols of certification that are entirely based on Gaussian operations and measurements, making them immediately suited to current technology implementation.

From the formal point of view, we notice that the MDI witnessing is based on the separation among different sets of induced measurements acting the input states, which in turn can be reconstructed via the statistics of the experiment. In particular, our protocol (as the one in~\cite{rosset2018resource}) can be seen to exploit the separation between the set of all POVMs and the set of one-way LOCC POVMs (cf. App.~\ref{sec:EB_bound_proof}). A similar argument can be applied to the case of entanglement witnessing~\cite{buscemi2012all,abiuso2021measurement}.
A general understanding, and a classification of the quantum certification protocols based on \emph{adversarial metrology}, is lacking and represents an interesting possible outlook of these works.

Finally, while tailored to the certification of quantum memories, we believe our protocol can be used in order to certify other types of non-EB operations, such as transmission lines, and optical transducers~\cite{zeuthen2020figures,lauk2020perspectives,bock2023calibration}.



%
%
%

\acknowledgements
\begin{wrapfigure}{R}[0cm]{2cm}
\includegraphics[width=2cm]{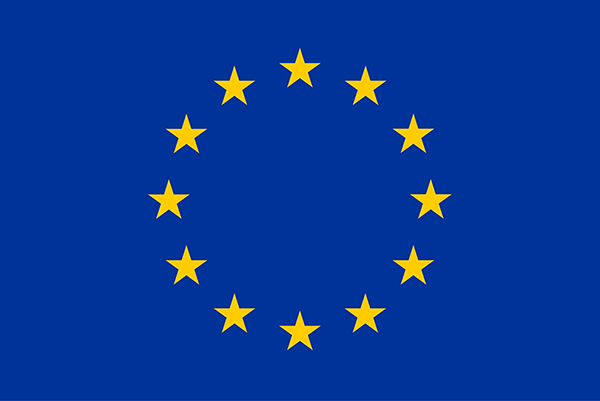}
\end{wrapfigure}
I thank Antonio Acín and Miguel Navascués for valuable feedback and discussion.
This project was funded within the
QuantERA II programme, that has received funding from the European Union’s Horizon 2020 research and innovation programme under Grant Agreement No 101017733, and from the Austrian Science Fund (FWF), project I-6004. Support is acknowledged by the Government of Spain (Severo Ochoa CEX2019-000910-S and NextGeneration EU PRTR-C17.I1), Fundació Cellex, Fundació Mir-Puig, Generalitat de Catalunya (CERCA program), European Union (QSNP, 101114043).



\medskip
\bibliography{MemoVer.bib}

\onecolumngrid
\appendix

\section{Entanglement breaking bound}
\label{sec:EB_bound_proof}
In this section we prove our first main result, i.e. the bound~\eqref{eq:EB_bound_main} in the main text.

\paragraph*{Notation.} 
We call $A$ ($B$) the Hilbert space of $\ket{\alpha}$ ($\ket{\beta}$).
Also, we will denote $\langle O \rangle$ the full average of any quantity $O$ on all stochastic variables, while the partial averages on the distributions of $\alpha$ and $\beta$ are denoted as
\begin{align}
    \langle O \rangle_\alpha:=\int \D \alpha\; P(\alpha)\; O\;,
    \quad
    \langle O \rangle_\beta:=\int \D \beta\; P(\beta)\; O\;,
    \quad\langle O \rangle_{\alpha,\beta}:= \int \D\alpha\D\beta\; P(\alpha)P(\beta) \;O \;. 
\end{align}

\paragraph*{Proof of~\eqref{eq:EB_bound_main} (Result~\ref{res1}).}
Eve is required to output two numbers, $\xi_x$ and $\xi_p$, in order to minimize the score represented by average error~\eqref{eq:score}
\begin{align}
\label{eqapp:wit}
    \langle\mathcal{W}\rangle= \langle\mathcal{W}_x\rangle+\langle\mathcal{W}_p\rangle
    :=\left\langle 
   \left(\xi_x - (\alpha_x+\beta_x)\right)^2
   \right\rangle 
   +
   \left\langle 
   \left(\xi_p - (\alpha_p-\beta_p)\right)^2
   \right\rangle  \;.
\end{align}
If the memory channel $\mathcal{M}$ (acting on the first input $\ket{\alpha}$) is EB, it can be expressed as a measure and prepare channel, therefore
\begin{align}
\label{eqapp:M_EB_Na}
    \mathcal{M}[\rho_A]=\int \D a \ \rho^{(a)}_{A'} \Tr[N^a_A \rho_A]\;,
\end{align}
where $N^a_A$ is a POVM.
This means that, upon receiving the states $\ketbra{\alpha}$ and $\ketbra{\beta}$, whatever estimation $\vec{\xi}:=\{\xi_x,\xi_p\}$ by Eve, it will be given by a measurement of the form
\begin{align}
\label{eq:POVM_EB}
    P(\vec{\xi}|\ket{\alpha},\ket{\beta}) 
    =\int \D a\; \Tr[M^{\vec{\xi}}_{A'B}(\rho^a_{A'}\otimes\ketbra{\beta}_B)]\;\Tr[N^{a}_A\ketbra{\alpha}_A]
    =   \int \D a\;\Tr[M^{\vec{\xi}|a}_B\ketbra{\beta}_B]\Tr[N^{a}_{A}\ketbra{\alpha}_A]
    \;,
\end{align}
where we defined $M^{\vec{\xi}|a}_B:=\Tr_A'[M^{\vec{\xi}}_{A'B}\rho^a_{A'}]$. This means that the output above can be expressed as a one-way LOCC bipartite measurement
\begin{align}
     P(\vec{\xi}|\ket{\alpha},\ket{\beta}) =\int \D a\; P(\vec{\xi}|a,\ket{\beta}) P(a|\ket{\alpha}) \;.
\end{align}
This decomposition is the fundamental ingredient in the proof of our bound. Consider now the average value of $\mathcal{W}_x$~\eqref{eqapp:wit}
\begin{align}
   \left\langle 
   \left(\xi_x - (\alpha_x+\beta_x)\right)^2
   \right\rangle 
   =
   \left\langle 
   \left(\xi_x - {\bar{\xi}_x}|_a + {\bar{\xi}_x}|_a - (\alpha_x+\beta_x)\right)^2
   \right\rangle \;,
\end{align}
where we introduced the average of $\xi_x$ for fixed $a$ (averaged also on $\beta$) i.e.
\begin{align}
   {\bar{\xi}_x}|_a := \int \D \xi_x \D\beta  \; \xi_x \;\Tr[M^{{\xi_x}|_a}_B\ketbra{\beta}_B]p(\beta)
\end{align}
This means that ${\bar{\xi}_x}|_{a}$ is a \emph{deterministic function} of $a$ (for the given POVM of Eve). Crucially, such function can be defined only thanks to the decomposition~\eqref{eq:POVM_EB}, i.e. thanks to the hypothesis of $\mathcal{M}$ being entanglement breaking.
Now it holds
\begin{align}
\label{eq:W_x_split}
    \langle\mathcal{W}_x\rangle =\left\langle 
   \left(\xi_x - {\bar{\xi}_x}|_a + {\bar{\xi}_x}|_a- \alpha_x-\beta_x\right)^2
   \right\rangle 
   =\left\langle 
   \left(\xi_x - {\bar{\xi}_x}|_a - \beta_x\right)^2\right\rangle +\left\langle\left( {\bar{\xi}_x}|_a - \alpha_x\right)^2
   \right\rangle \;,
\end{align}
which can be seen to be true because the cross-term vanishes fixing $a$ and $\alpha$, and taking the partial average on $\xi$ and $\beta$ (notice also that $\beta$ is uncorrelated with $\{a,\alpha\}$). In fact,
\begin{align}
   \left\langle 
   \left(\xi_x - {\bar{\xi}_x}|_a -\beta_x\right)\left( {\bar{\xi}_x}|_a - \alpha_x\right)
   \right\rangle  = \left\langle 
   \left({\bar{\xi}_x}|_a - {\bar{\xi}_x}|_a - \langle\beta_x\rangle_\beta\right)\left( {\bar{\xi}_x}|_a - \alpha_x\right)
   \right\rangle_{a,\alpha} \;.
\end{align}
Without loss of generality we consider the distribution of $\beta$ to be centered in zero, so that the above cross term is null, and one is left with Eq.~\eqref{eq:W_x_split}. 

To conclude, it is enough now to notice that the two terms in~\eqref{eq:W_x_split} correspond to errors for quantum estimators of $\beta_x$ and $\alpha_x$ respectively. That is,
${\bar{\xi}_x}|_a $ is a function that depends solely on the random variable $a$, which is the output of a POVM on $\ket{\alpha}$.
The same is valid for the term $\xi_x - {\bar{\xi}_x}|_a - \beta_x$, which does not depend on $\ket{\alpha}$ after conditioning on $a$, and whose average can be computed by first fixing $a$. Once $a$ is fixed, the term $\xi_x - {\bar{\xi}_x}|_a$ is simply the output of a POVM on $\ket{\beta}$.
This means that the whole witness $\langle\mathcal{W}_x\rangle+\langle\mathcal{W}_p\rangle$, in the EB case, can be formally lower-bounded by standard Cramér-Rao bounds on displacement estimation~\cite{yuen1973multiple,genoni2013optimal,morelli2021bayesian}. 

What follows is the formalization of the above observations. One has, from the above manipulations in the EB case,
\begin{align}
    \langle\mathcal{W}_x + \mathcal{W}_p\rangle= \underbrace{
    \left\langle 
   \left(\xi_x - {\bar{\xi}_x}|_a -\beta_x\right)^2 +
   \left(\xi_p - {\bar{\xi}_p}|_a - \beta_p\right)^2 
   \right\rangle
   }_{\langle\mathcal{W}^{(B)}\rangle}
   +
   \underbrace{
   \left\langle\left( {\bar{\xi}_x}|_a - \alpha_x\right)^2
   +
   \left( {\bar{\xi}_p}|_a - \alpha_p\right)^2
   \right\rangle 
   }_{\langle\mathcal{W}^{(A)}\rangle}\;.
\end{align}
We treat the two terms separately. For $\mathcal{W}^{(A)}$. Notice that, as explained above $\vec{\bar{\xi}}|_a=\{\bar{\xi}_x|_a,\bar{\xi}_p|_a\}$ is a deterministic function of $a$, and therefore can be seen as the result of the POVM $N^{a}$ in~\eqref{eqapp:M_EB_Na}
\begin{align}
    P(\vec{\bar{\xi}}|_a|\ket{\alpha})=\Tr[N^{a}\ketbra{\alpha}]\;.
\end{align}
Secondly, for $\mathcal{W}^{(B)}$ one has
\begin{align}
    \langle \mathcal{W}^{(B)}\rangle\geq \min_a \langle  \mathcal{W}^{(B)}\rangle|_a\;.
\end{align}
However, for fixed $a$, the output $\vec{\xi}'|_a\equiv\vec{\xi}-\vec{\bar{\xi}}|_a$ is the output of a POVM acting on $\ket{\beta}$ alone, i.e. from~\eqref{eq:POVM_EB}
\begin{align}
    P(\vec{\xi}'|_a|\ket{\beta})=\Tr[M^{(\vec{\xi}'+\vec{\bar{\xi}}|_a)|_a}\ketbra{\beta}]\;.
\end{align}
As both $\langle\mathcal{W}^{(A)}\rangle$ and $\langle\mathcal{W}^{(B)}\rangle |_a$ correspond to average estimation errors locally measured on $\ket{\alpha}$ and $\ket{\beta}$, one can therefore use Cramér-Rao bounds to lower bound them. 
As we are in an adversarial scenario, we allow Eve to have prior knowledge of the input distributions $P(\alpha)$, $P(\beta)$, which she might reconstruct after sufficiently many rounds of the experiment. 
In particular, assuming Gaussian input distributions
\begin{align}
\label{eqapp:gauss_prior}
    P(\alpha)=\frac{e^{-|\alpha|^2/\sigma^2_\alpha}}{\sigma^2_\alpha\pi}\;,\quad P(\beta)=\frac{e^{-|\beta|^2/\sigma^2_\beta}}{\sigma^2_\beta\pi}\;,
\end{align}
it follows, from standard displacement estimation theory~\cite{yuen1973multiple,genoni2013optimal}, that
\begin{align}
\label{eqapp:WA_lb}
    \langle\mathcal{W}^{(A)}\rangle &\geq \frac{\sigma_\alpha^2}{1+\sigma_\alpha^2}\;, \\
    \langle\mathcal{W}^{(B)}\rangle|_{a} &\geq \frac{\sigma_\beta^2}{1+\sigma_\beta^2} \;.
\label{eqapp:WB_lb}
\end{align}
Our main bound~\eqref{eq:EB_bound_main} therefore follows straightforwardly from $\langle\mathcal{W}\rangle=\langle\mathcal{W}^{(A)}+\mathcal{W}^{(B)}\rangle\geq \langle\mathcal{W}^{(A)}\rangle+\min_a\langle\mathcal{W}^{(B)}\rangle|_a$, which concludes the proof.

\subsection{Using non-Gaussian priors}
\label{app:non-gaussian_priors}
We notice that the bounds~\eqref{eqapp:WA_lb}-\eqref{eqapp:WB_lb} above, are derived from the multi-parameter quantum Cramér-Rao bound based on the right logarithmic derivative (RLD)~\cite{holevo2011probabilistic}, in its Bayesian form with additional prior information~\cite{yuen1973multiple}. This bound, when applied to the minimization of the sum of the single parameter variances (in our case $\alpha_x$, $\alpha_p$, and similarly $\beta_x$, $\beta_p$), is presented e.g. in Eq.~(14) of Ref.~\cite{genoni2013optimal}. The information associated to the prior distribution $P(\alpha)\equiv P(\alpha_x,\alpha_p)$ is encoded in the prior's Fisher Information Matrix (FIM) defined in Eq.~(10) of the same Ref.~\cite{genoni2013optimal}, that is
\begin{align}
    F_{ij}:=\int \D\alpha_x \D\alpha_p\; P(\alpha_x,\alpha_p)\frac{\partial \log P(\alpha_x,\alpha_p)}{\partial \alpha_i} \frac{\partial \log P(\alpha_x,\alpha_p)}{\partial \alpha_j}\;,
\end{align}
where $i$ and $j$ can be equal to $x$ and $p$. For the Gaussian distribution~\eqref{eqapp:gauss_prior} used to derive the entanglement breaking bound above, one has $P(\alpha)=\frac{1}{\pi\sigma^2}\exp{[-(\alpha_x^2+\alpha_p^2)/\sigma^2]}$ and simple algebra yields a diagonal $F$
\begin{align}
\label{eq:gauss_fim}
    F^{\rm Gauss}=\begin{pmatrix}
    \frac{2}{\sigma^2} & 0 \\
    0 & \frac{2}{\sigma^2}
    \end{pmatrix}\;.
\end{align}
The property of $F$ being diagonal clearly holds for any distribution in the form $P(\alpha_x,\alpha_p)=P(\alpha_x)P(\alpha_p)$ which is factorized. In such a case $F_{ij}=f_{i}\delta_{ij}$, where $f_i$ is the single parameter Fisher information relative to $P(\alpha_i)$,
\begin{align}
    F^{\rm symm}=
    \begin{pmatrix}
    f_x & 0 \\
    0 & f_p
    \end{pmatrix}\;,\qquad f_i=\int \D\alpha_i\; P(\alpha_i)\left(\frac{\partial \log P(\alpha_i)}{\partial \alpha_i}\right)^2\;.
\end{align}
As an example, it is possible to choose an almost-flat distribution on a square of size $l$, defined as
\begin{align}
    P(\alpha_x,\alpha_p)=\frac{1}{l^2}I_{\delta,l}(\alpha_x)I_{\delta,l}(\alpha_p)
\end{align}
where $I_{\delta,l}$ is a smooth version of the indicator function on the interval $[-l/2,+l/2]$, 
\begin{align}
  I_{\delta,l}=\begin{cases}
  0 & x\leq -\frac{l}{2}-\frac{\delta}{2} \\
  \frac{1}{2}+\frac{1}{2}\sin(\frac{\pi}{\delta}(x+\frac{l}{2})) & -\frac{l}{2}-\frac{\delta}{2}\leq x \leq -\frac{l}{2}+\frac{\delta}{2}\\
  1 & -\frac{l}{2}+\frac{\delta}{2}\leq x \leq \frac{l}{2}-\frac{\delta}{2}\\
   \frac{1}{2}-\frac{1}{2}\sin(\frac{\pi}{\delta}(x-\frac{l}{2})) & \frac{l}{2}-\frac{\delta}{2}\leq x \leq \frac{l}{2}+\frac{\delta}{2}\\
   0 & x\geq \frac{l}{2}+\frac{\delta}{2}
  \end{cases}
\end{align}
Note that for consistency it is needed $\delta\leq l$, and the limit cases $\delta=0$ and $\delta=l$, correspond respectively to $I_{\delta,l}$ being the indicator function and $I_{\delta,l}$ being a single symmetric cosinusoidal wave between $-l$ and $+l$.
For such a choice of $P(\alpha_x,\alpha_p)$ the corresponding fisher information matrix is easily computed as
\begin{align}
    F^{(\delta,l)}=
    \begin{pmatrix}
    \frac{\pi^2}{l\delta} & 0 \\
    0 & \frac{\pi^2}{l\delta}
    \end{pmatrix}\;.
\end{align}
We see that in the limit $\delta\rightarrow 0$ the FIM diverges and becomes useless for computing Cramér-Rao bounds, which become trivial in such limit. This is related to the fact that the multi-parameter Cram\'er Rao bounds are not tight in general \cite{holevo2011probabilistic} (notice that in case of Gaussian prior it can be saturated \cite{genoni2013optimal}). In general we see that the associated prior FIM becomes equivalent to the Gaussian case~\eqref{eq:gauss_fim}, provided
\begin{align}
    \frac{\pi^2}{l\delta}=\frac{2}{\sigma^2}\;.
\end{align}
In the limit $\delta=l$, this translates to
\begin{align}
   \frac{l}{\pi}=\frac{\sigma}{\sqrt{2}} \;.
\end{align}

\section{Incompatible measurements, incompatibility breaking, and entanglement breaking channels}
\label{app:IB_EB_etc}

Not all measurements can be performed simultaneously on a quantum system. Non-commuting observables cannot be measured at the same time.
In general, consider set of measurements (or their induced observables), indexed by $y\in\mathcal{Y}$, with outcomes $\xi$, therefore represented by POVM operators $N^{\xi|y}$,
\begin{align}
\label{eqapp:comp_meas}
    N^{\xi|y}\geq 0\;,\quad  \int\D \xi\;  N^{\xi|y}=\eye \quad \forall y\;.
\end{align}
The measurements indexed by $y$ are defined to be jointly measurable, or simply \emph{compatible}, if it exists some POVM $M^{a}$ and conditional probability $P(\xi|a,y)$ such that
\begin{align}
    N^{\xi|y}=\int \D a\; P(\xi|a,y) M^{a}\;.
\end{align}
Operationally speaking, the above equation means that any measurement in $\mathcal{Y}$ can be simulated by performing the same, parent-measurement $M$, and post-processing classically the output via $P(\xi|a,y)$. 

Incompatibility breaking (IB) channels~\cite{heinosaari2015incompatibility} are defined as those that degrade quantum states enough to make a set of previously incompatible measurements, compatible. More precisely, consider a CPTP map $\Lambda[\rho]$ acting on quantum states. Given the dual channel $\Lambda^\dagger$ in the Heisenberg picture
\begin{align}
    \Tr[O\Lambda[\rho]]=\Tr[\Lambda^\dagger[O]\rho]\quad \forall O\geq 0, \rho \geq 0\;,
\end{align}
then $\Lambda$ is said to break the incompatibility of a set of measurements $N^{\xi|y}$ if
\begin{align}
    N'^{\xi|y}=\Lambda^\dagger[N^{\xi|y}] \quad \text{is compatible for } y\in\mathcal{Y}\;.
\end{align}
This corresponds to the fact that, once the noisy action of $\Lambda$ has taken place, all the possible observables $N^{\xi|y}$, correspond to observables that could have been jointly measured in the absence of noise, as $\Tr[N^{\xi|y}\Lambda[\rho]]=\Tr[N'^{\xi|y}\rho]$ by definition.

Notice that the notion of incompatibility breaking depends on the class of chosen incompatible observables. With a small abuse of notation, let us indicate with $\mathcal{Y}$-IB the set of channels that break the incompatibility of all observables in $\mathcal{Y}$. It follows from the definition that
\begin{align}
    \mathcal{Y}_1\subseteq \mathcal{Y}_2 \; \Rightarrow \; \mathcal{Y}_2\text{-IB }\subseteq \mathcal{Y}_1\text{-IB }.
\end{align}
It is then possible to define the set of channels that break the incompatibility of any set of observables, and simply label them IB channels,
\begin{align}
    \text{IB } := \bigcap_\mathcal{Y} \mathcal{Y}\text{-IB } \;,
\end{align}
for all possible set of measurements $\mathcal{Y}$. 
Notably, EB channels are known to be a (strict) subset of IB channels~\cite{heinosaari2015incompatibility}. This is due to the fact that EB channels correspond to measure and prepare channels, i.e.
\begin{align}
    \mathcal{E}[\rho]=\int \D a\; \rho^{(a)}\Tr[N^{a}\rho]\;,
\end{align}
from which it follows
\begin{align}
     \mathcal{E}^\dagger[N^{\xi|y}]=\int \D a\; N^{a}\Tr[N^{\xi|y}\rho^{(a)}]\;,
\end{align}
which is exactly in the form~\eqref{eqapp:comp_meas}, independently of the original POVM $N^{\xi|y}$. It follows that
\begin{align}
    \text{EB}\subseteq \text{IB}\;.
\end{align}
The existence of non-EB channels that are IB was proven in~\cite{heinosaari2015incompatibility}.

Finally, in~\cite{heinosaari2015breaking} the authors studied the case of continuous variables, and in particular they characterized Gaussian incompatibility breaking, i.e. channels that break the incompatibility of measurements having a Gaussian Wigner distribution, such as standard homodyne measurements (the Wigner function of the quadratures $\hat{x}$ and $\hat{p}$ can be thought of Gaussian functions with zero variance, i.e. delta functions). These are called \emph{Gaussian incompatibility breaking} (gIB).
We discuss in the following section properties of Gaussian channels that are EB, IB, or gIB, and their characterization (see~\cite{heinosaari2015breaking}). For the moment, it is enough to notice that also when restricting to Gaussian channels only, the following hierarchy trivially holds
\begin{align}
    \text{EB}\subset \text{IB} \subset \text{gIB}\;.
\end{align}

\section{Witnessing non Gaussian incompatibility breaking channels with homodyne measurements}
\label{sec:gauss_app}

In this Appendix, we prove that the bound~\eqref{eq:EB_bound_main} can be violated with any memory $\mathcal{M}$ corresponding to a Gaussian channel that is not Gaussian incompatibility breaking (non-gIB)~\cite{heinosaari2015breaking}. That is, if Eve possesses a memory $\M$ belonging to such class, she can append Gaussian operations to the memory and perform a simple protocol that certifies $\mathcal{M}$.

We start with a quick introduction to Gaussian states and Gaussian channels, as well as review of criteria of complete positivity, entanglement breaking-ness, and Gaussian incompatibility breaking-ness for such channels.
After setting the necessary tools and terminology, we provide in Sec.~\ref{app:proof_res2} the proof of Result~\ref{res2} of the main text.

\subsection{Gaussian states and channels} 
\label{app:gauss_channels}
Gaussian states are those that have a Gaussian characteristic function, or equivalently a Gaussian Wigner distribution~\cite{ferraro2005gaussian,adesso2014continuous}. They can be fully described via their covariance matrix and displacement vector. More precisely, for an $n$-mode state, one introduces the $2n$-dimensional vector of canonical quadrature observables
\begin{align}
    \vec{\hat{r}}:=\begin{pmatrix}
        \vec{\hat{x}}\\ \vec{\hat{p}}
    \end{pmatrix}\;,
\end{align}
Where $\hat{x}_i=\frac{a_i+a_i^\dagger}{\sqrt{2}}$ is the $n$-dimensional vector of $\hat{x}$ quadratures for each mode, and similarly $\hat{p}_i=\frac{a_i-a_i^\dagger}{i\sqrt{2}}$ .
One can then define the $2n\times 2n$ covariance matrix and the $2n$ displacement vector of any state as
\begin{align}
    \Sigma_{ij}:=\langle \{\Delta\hat{r}_i,\Delta\hat{r}_j\}\rangle\;, \quad D_i=\langle\hat{r}_i\rangle\;,
\end{align}
where $\Delta\hat{r}_i:=\hat{r}_i-\langle\hat{r}_i\rangle$. Gaussian quantum states are completely characterized by their corresponding $\Sigma$ and $\vec{D}$.
The \emph{bona fide} condition for a given $\Sigma$ to represent a proper quantum state is given by
\begin{align}
    \Sigma+i\Omega\geq 0\;,\quad \Omega:=
    \begin{pmatrix}
    0 & -\eye \\ \eye & 0
    \end{pmatrix}\;,
\end{align}
which in turns follows from the commutation relations $[\hat{x},\hat{p}]=1$ and the matrix $\langle \Delta\hat{r}_i \Delta\hat{r}_j\rangle$ being positive.

For a $1$-mode coherent state $\ket{\alpha}$, one has $\Sigma=\eye$ and $\vec{D}=\sqrt{2}(\alpha_x,\alpha_p)$.

Quantum channels that preserves Gaussianity of the states are called Gaussian channels~\cite{adesso2014continuous,eisert2005gaussian} and can be represented from their action on $\Sigma$ and $\vec{D}$. In its most general form, a $n$-mode Gaussian channel $\mathcal{G}(K,M,\vec{c})$ operates as 
\begin{align}
\label{eqapp:gauss_chann_action1}
    \Sigma &\rightarrow K\Sigma K^\intercal + M \;,\\
\label{eqapp:gauss_chann_action2}
    \vec{D} &\rightarrow K \vec{D} + \vec{c}\;.
\end{align}
Here $K$ is a real $2n\times 2n$ matrix, $M$ is a real positive $2n\times 2n$ matrix, and $\vec{c}$ a $2n$ real vector.
As the triple $\{K,M,\vec{c}\}$ fully specifies $\mathcal{G}$, it is possible to give characterizations of different sets of Gaussian superoperators via conditions on the element of the triple. Notice that a nonzero $\Vec{c}$ corresponds to a simple unitary displacement and it is typically irrelevant for properties such as channel positivity, incompatibility breaking, entanglement breaking, and others. In the following we summarize three main criteria for Gaussian superoperators that are relevant to our work. The proof of these can be found in the references~\cite{adesso2014continuous,eisert2005gaussian,holevo2008entanglement-breaking,heinosaari2015breaking}.

\vspace{0.5cm}
\paragraph*{Matrix Criteria for Gaussian Channels.}

\begin{itemize}
    \item{\bf{CP condition.}} First, a Gaussian superoperator $\mathcal{G}(K,M,\vec{c})$ acting on states as~\eqref{eqapp:gauss_chann_action1}-\eqref{eqapp:gauss_chann_action2}, is a proper quantum channel, i.e. a completely positive (CP) map if and only if the following condition holds
    \begin{align}
    \label{eqapp:cond_CP}
        M+i\Omega-iK\Omega K^\intercal \geq 0\;.
    \end{align}
    \item{\bf{gIB condition.}} Secondly, a Gaussian channel $\mathcal{G}(K,M,\vec{c})$ is Gaussian incompatibility breaking (gIB) if and only if
    \begin{align}
    \label{eqapp:cond_gIB}
        M-iK\Omega K^\intercal  \geq 0\;.
    \end{align}
    \item{\bf{EB condition.}} Finally, a Gaussian channel $\mathcal{G}(K,M,\vec{c})$ is entanglement breaking (EB) if and only if
    \begin{align}
    \nonumber
        &M =M_1+M_2 \quad \text{with}\\ 
        &M_1+i\Omega \geq 0\quad \wedge \quad M_2-iK \Omega K^\intercal  \geq 0\;.
         \label{eqapp:cond_EB}
    \end{align}
\end{itemize}

The above criteria take a simpler form when one restricts to the one-mode scenario ($n=1$), in which case $M$ and $K$ are $2\times 2$ matrices, as well as $\Omega=\begin{pmatrix}
    0 & -1\\ 1 & 0
    \end{pmatrix}$.
In fact, in such case
\begin{align}
    K\Omega K^\intercal=\Omega \det{K} \quad \forall K\;.
\end{align}
Moreover given any real positive $2\times 2$ matrix $M$ and real scalar $\kappa$, simple algebra shows that
\begin{align}
    M+i \kappa \Omega \geq 0 \Leftrightarrow \det{M}\geq \kappa^2\;.
\end{align}
The above two observations imply that the criteria above can be remapped as follows (cf.~\cite{siudzinska2019geometry}).

\paragraph*{Matrix Criteria for 1-Mode Gaussian Channels.}
\begin{itemize}
    \item{\bf{1-mode CP condition.}} A 1-mode Gaussian map $\mathcal{G}(K,M,\vec{c})$ is CP if and only if $M\geq 0$ and
    \begin{align}
      \label{eqapp:cond_CP_1mode}
        \det M\geq (\det K -1)^2\;.
    \end{align}
    \item{\bf{1-mode gIB condition.}} A 1-mode Gaussian channel $\mathcal{G}(K,M,\vec{c})$ is gIB if and only if $M\geq 0$ and
    \begin{align}
     \label{eqapp:cond_gIB_1mode}
        \det M\geq \det K ^2\;.
    \end{align}
    \item{\bf{1-mode EB condition.}} A 1-mode Gaussian channel $\mathcal{G}(K,M,\vec{c})$ is EB if and only if $M\geq 0$ and
    \begin{align}
     \label{eqapp:cond_EB_1mode}
    \det M\geq (\det K +1)^2\;.
    \end{align}
\end{itemize}

\paragraph*{Photon-loss example.}
We analyze here the photon-loss dynamics considered in the main text~\eqref{eq:PL_quadrature_heis}, which can be modelled as a beamsplitter with transmissivity $\eta$ mixing the original state with a vacuum state. That is, in the  Heisenberg picture the annihilation operator transforms as
\begin{align}
    \hat{a}'=\sqrt{\eta}\hat{a}+\sqrt{1-\eta}\hat{a}^{(0)}\ , 
\end{align}
where $\hat{a}^{(0)}$ is the annihilation operator acting on the vacuum, and $0\leq \eta\leq 1$. Such channel is Gaussian and corresponds to
\begin{align}
   M=(1-\eta)\eye\;, K=\sqrt{\eta}\eye\;, \vec{c}=0\;.
\end{align}
Therefore, analyzing the conditions~(\ref{eqapp:cond_CP_1mode}-\ref{eqapp:cond_EB_1mode}) it easily follows that such channel is CP for all values of $\eta\in[0,1]$, it is gIB for $\eta\in[0,0.5]$, and EB only in the extremal erasure case $\eta=0$ (cf. Fig.~\ref{fig:eta_photonloss} in main text).

\subsection{Violating the bound~\eqref{eq:EB_bound_main} with non-gIB channels}
\label{app:proof_res2}
Here we prove our Result~\ref{res2} of the main text, which can be formulated as follows: assume that Eve possesses a memory $\mathcal{M}$ consisting in a Gaussian channel that is not Gaussian incompatibility breaking (non-gIB). Then, there exist two Gaussian ``recalibration" channels $\mathcal{G}_{1,2}$ such that i.e. the witness~\eqref{eq:EB_bound_main} can be violated by achieving a value strictly smaller than $2$, using $\mathcal{M}'=\mathcal{G}_2\circ\mathcal{M}\circ\mathcal{G}_1$ in the protocol described in the main text (cf. Fig.~\ref{fig:setup}). That is, explicitly, Eve mixes the two states $\rho_{\alpha'}$ and $\rho_\beta$
\begin{align}
    \rho_{\alpha'}&=\mathcal{M}'[\ketbra{\alpha}]\equiv \mathcal{G}_2\circ\mathcal{M}\circ\mathcal{G}_1[\ketbra{\alpha}] \;,\\ 
    \rho_{\beta}&=\ketbra{\beta}\;,
\end{align}
in a 50:50 beamsplitter, after which she measures $\hat{x}$ and $\hat{p}$. Define
\begin{align}
    \{\Sigma^{(\alpha')},  \vec{D}^{(\alpha')}\}:= \text{the covariance matrix and displacement vector of }  \rho_{\alpha'}\;,
\end{align}
and similarly for $\rho_\beta=\ketbra{\beta}$. The latter is a coherent state and satisfies 
\begin{align}
    \Sigma^{(\beta)}={\eye}, \; \vec{D}^{(\beta)}=\sqrt{2}(\beta_x,\beta_p)\;.
\end{align}
Assume now that 
\begin{align}
\label{eqapp:assumption}
    \emph{Assumption: } \vec{D}^{(\alpha')}=\sqrt{2}(\alpha_x,\alpha_p)\;.
\end{align}
If this is the case, it follows that the measurements of Eve are unbiased, that is, she measures
\begin{align}
\hat{x}_E:=\frac{\hat{x}_{\alpha'}+\hat{x}_\beta}{\sqrt{2}}\;,\quad
\hat{p}_E:=\frac{\hat{p}_{\alpha'}-\hat{p}_\beta}{\sqrt{2}}\;,
\end{align}
and the above assumption is reflected in
\begin{align}
    \langle\hat{x}_E\rangle =\alpha_x+\beta_x \;,\quad  \langle\hat{p}_E\rangle =\alpha_p-\beta_p\;.
\end{align}
When the above unbiased-ness property is satisfied, it follows that regardless of the distributions $P(\alpha)$, $P(\beta)$, that are used to sample the inputs, one has that the average value of witness~\eqref{eq:score} is given by
\begin{align}
\nonumber
\langle\mathcal{W}\rangle &=    {\rm Var}[\hat{x}_E] + {\rm Var}[\hat{p}_E]= \frac{1}{2} \left(
{\rm Var}[\hat{x}_{\alpha'}]+{\rm Var}[\hat{p}_{\alpha'}]+
{\rm Var}[\hat{x}_{\beta}]+{\rm Var}[\hat{p}_{\beta}]
\right) \\
&=\frac{1}{4}\Tr[\Sigma^{(\alpha')}+ {\Sigma^{(\beta)}}]\;.
\end{align}
Given that $\Sigma^{(\beta)}=\eye$, it follows that $\Tr[\Sigma^{(\beta)}]=2$ and the whole witness violation is equivalent to
\begin{align}
\label{eqapp:W<2_alpha}
    \langle\mathcal{W}\rangle < 2  \Leftrightarrow \Tr[ \Sigma^{(\alpha')}]<6\;.
\end{align}
We are now ready to prove Result~\ref{res2}. That is, we prove that once a non-gIB $\mathcal{M}$ is given, it is possible to find $\mathcal{G}_1$ and $\mathcal{G}_2$ such that the assumption~\eqref{eqapp:assumption} holds and and $\Tr [\Sigma^{(\alpha')}]<6$.

\paragraph*{Proof.}
Consider the action of $\mathcal{M}$
\begin{align}
\mathcal{M}&=\mathcal{G}(K,M,\vec{c})\;,\\
    \Sigma &\rightarrow K\Sigma K^\intercal + M \;,\\
    \vec{D} &\rightarrow K\vec{D} +\vec{c}\equiv  K\vec{D}  \;.
\end{align}
Here we assumed, without loss of generality, that $\vec{c}=0$, as it consists in a unitary displacement which can always be reverted via $\mathcal{G}(\eye,0,-\vec{c})$. 

The input coherent state is characterized by $\Sigma^{(\alpha)}={\eye}$ and $\Vec{D}^{(\alpha)}=\sqrt{2}(\alpha_x,\alpha_p)$.
On such input acts the composition of the three operations $\mathcal{G}_2\circ\mathcal{M}\circ\mathcal{G}_1$. Assume the channels $\mathcal{G}_{1,2}$ to be of the form
\begin{align}
    \mathcal{G}_1:=\mathcal{G}(K_1,M_1,0)\;,\\
    \mathcal{G}_2:=\mathcal{G}(K_2,M_2,0)\;.
\end{align}
From this it follows that the action of these operations on $\{\Sigma,\Vec{D}\}$ is
\begin{align}
\label{eqapp:tranformation_steps}
    \Sigma &\rightarrow K_1\Sigma K_1^{\intercal} + M_1 & & \rightarrow  K (K_1\Sigma K_1^{\intercal} + M_1) K^\intercal + M  & & \rightarrow  K_2(K (K_1\Sigma K_1^{\intercal} + M_1) K^\intercal + M)K_2^\intercal + M_2\;, \\
    \vec{D} & \rightarrow K_1\Vec{D} & & \rightarrow KK_1\Vec{D} & & \rightarrow K_2KK_1\Vec{D}\;.
\end{align}

Consider now the $2\times 2$ matrix $K^{-1}MK^{-1\intercal}$, which is manifestly positive. Williamson theorem then guarantees that it can be symplectically diagonalized, i.e.
\begin{align}
\label{eqapp:symp_diag}
 K^{-1}MK^{-1\intercal}\equiv  S^{-1}\lambda
\eye S^{-1\intercal}\;, \quad \lambda=\sqrt{\frac{\det{M}}{\det K^2}}\;,
\end{align}
for some symplectic matrix $S$ satisfying $S\Omega S^\intercal=\Omega$. Importantly, thanks to the non-gIB condition~\eqref{eqapp:cond_gIB_1mode} of $\mathcal{M}$, we have
\begin{align}
\label{eqapp:lambda<1}
    0\leq \lambda < 1\;.
\end{align}
We choose then $\mathcal{G}_1$ to be defined by 
\begin{align}
K_1=S^{-1}\;,\quad M_1=0\;.
\end{align}
This is a valid Gaussian CP channel as it satisfies the condition~\eqref{eqapp:cond_CP}-\eqref{eqapp:cond_CP_1mode}, and in the first step of~\eqref{eqapp:tranformation_steps} we obtain for the covariance matrix
\begin{align}
     \Sigma^{(\alpha)}\equiv {\eye} \rightarrow S^{-1} \Sigma^{(\alpha)}S^{-1\intercal}=S^{-1}S^{-1\intercal}  \;.
 \end{align}
 Secondly, the action of $\mathcal{M}$ applies, and the covariance matrix becomes
 \begin{align}
      S^{-1}S^{-1\intercal} \rightarrow  K S^{-1}S^{-1\intercal} K^\intercal + M\;.
 \end{align}
Finally, the action of $\mathcal{G}_2$. We choose 
\begin{align}
K_2=S K^{-1}\;, 
\end{align}  
and obtain after the action of $\mathcal{G}_2$,
\begin{align}
     K S^{-1}S^{-1\intercal} K^\intercal + M \rightarrow {\eye}+ S K^{-1}MK^{-1\intercal}S^\intercal + M_2 = (1+\lambda)\eye + M_2=:\Sigma^{(\alpha')}\;,
\end{align}
where we used the symplectic diagonalization~\eqref{eqapp:symp_diag}.
This corresponds to the final covariance matrix $\Sigma^{(\alpha')}$.
We therefore want to prove that it is possible to have~\eqref{eqapp:W<2_alpha}
\begin{align}
    \Tr[(1+\lambda)\eye+M_2]<6\;.
\end{align}
Thanks to the fact that $\mathcal{M}$ is non-gIB we have $\lambda<1$ (see~\eqref{eqapp:lambda<1}). It follows that the above inequality is satisfied if we can admit $M_2$ to have a trace
\begin{align}
    \Tr[M_2]\leq 2\;.
\end{align}
The only constraint the $M_2$ has to satisfy is that, together with $K_2\equiv K^{-1}$, it has to form a proper Gaussian channel, i.e. it needs to satisfy the CP condition~\eqref{eqapp:cond_CP_1mode}
\begin{align}
    \det M_2\geq (\det K_2-1)^2=(\det K^{-1}-1)^2\;.
\end{align}
To bound the determinant of $K$, notice that by hypothesis $\mathcal{M}$ is CP and non-gIB. This means, from~\eqref{eqapp:cond_CP_1mode} and~\eqref{eqapp:cond_gIB_1mode}, that
\begin{align}
    \det K^2>\det M\geq (\det K -1)^2\;,
\end{align}
which implies straightforwardly
\begin{align}
\det K > \frac{1}{2}\;.    
\end{align}
Then, we have
\begin{align}
  -1 \leq \det K^{-1}-1 < 1\;.
\end{align}
It follows that by choosing
\begin{align}
    \det M_2=1
\end{align}
the map $\mathcal{G}_2=\mathcal{G}(K^{-1},M_2,0)$ is guaranteed to be a proper Gaussian CP channel.
Under this constrain, the minimum trace of $M_2$ is obtained for the choice 
\begin{align}
    M_2=\eye\;,
\end{align}
which satisfies $\Tr[M_2]=2$, and therefore
\begin{align}
    \Tr[\Sigma^{(\alpha')}]<6\;,
\end{align}
as desired to prove our result.

Finally, notice that the assumption~\eqref{eqapp:assumption} of $\Vec{D}^{(\alpha')}=\vec{D}^{(\alpha)}=\sqrt{2}(\alpha_x,\alpha_p)$ is automatically satisfied as 
\begin{align}
   \Vec{D}^{(\alpha')}=K_2KK_1\vec{D}^{(\alpha)}\;,\quad \text{given that we chose } K_1=S^{-1} \text{ and } K_2=S K^{-1}\;.
\end{align}
This completes the proof.

\end{document}